\documentclass[prb,twocolumn,showpacs,superscriptaddress]{revtex4}

\usepackage{graphicx}
\usepackage{dcolumn}
\usepackage{amsmath}
\usepackage{color}
\usepackage[english]{babel}

\newcommand{\RbFeSeSe}{Rb$_{0.74}$Fe$_{1.6}$Se$_2$}

\begin{document}

\title{Tuning orbital-selective correlations in superconducting Rb$_{0.75}$Fe$_{1.6}$Se$_{2-z}$S$_z$}

\author{Zhe~Wang}
\affiliation{Experimental Physics V, Center for Electronic
Correlations and Magnetism, Institute of Physics, University of Augsburg, D-86135 Augsburg, Germany}

\author{V.~Tsurkan}
\affiliation{Experimental Physics V, Center for Electronic
Correlations and Magnetism, Institute of Physics, University of Augsburg, D-86135 Augsburg, Germany}
\affiliation{Institute of Applied Physics, Academy of Sciences of Moldova, MD-2028 Chisinau, Republic of Moldova}

\author{M.~Schmidt}
\affiliation{Experimental Physics V, Center for Electronic
Correlations and Magnetism, Institute of Physics, University of Augsburg, D-86135 Augsburg, Germany}

\author{A.~Loidl}
\author{J.~Deisenhofer}
\affiliation{Experimental Physics V, Center for Electronic
Correlations and Magnetism, Institute of Physics, University of Augsburg, D-86135 Augsburg, Germany}

\date{\today}

\begin{abstract}
We report on terahertz time-domain spectroscopy on superconducting and metallic iron chalcogenides Rb$_{0.75}$Fe$_{1.6}$Se$_{2-z}$S$_z$.
The superconducting transition is reduced from $T_c=32$~K ($z=0$) to 22~K ($z=1.0$),
and finally suppressed ($z=1.4$) by  isoelectronic substitution of Se with S.
Dielectric constant and optical conductivity exhibit a metal-to-insulator transition associated with an orbital-selective Mott phase.
This orbital-selective Mott transition appears at higher temperature $T_{met}$ with increasing sulfur content, identifying sulfur substitution as an efficient parameter to tune orbital-dependent correlation effects in iron-chalcogenide superconductors.
The reduced correlations of the $d_{xy}$ charge carriers can account for the suppression of the superconductivity and the pseudogap-like feature between $T_c$ and $T_{met}$ that was observed for $z=0$.
\end{abstract}


\maketitle

\section{Introduction}

The concept of orbital differentiation has been suggested to offer a common ground to understand high-temperature superconductivity in iron-based superconductors and in the cuprates.\cite{Medici14,Misawa14,YuSong15} In the multi-orbital Fe-based superconductors orbital-dependent correlation effects have been predicted,\cite{Georges13,Yu13}
where the quasiparticles in one band can, for example, undergo a metal-insulator transition, while the other bands retain their metallic character.
This scenario has been called orbital-selective Mottness and the experimental observations in the \emph{A}$_{1-x}$Fe$_{2-y}$Se$_2$ family of iron-selenide superconductors with alkali metals \emph{A} = K, Rb, Cs make these materials representative models for tuning orbital-dependent correlation effects:
The occurrence of an orbital-selective crossover regime between metallic and insulating behaviors has been reported by angular-resolved photoemission spectroscopy (ARPES) for quasiparticles with $d_{xy}$ character as a function of temperature.\cite{Yi13,Yi15}
Using terahertz spectroscopy the orbital selective metal-insulator transition in superconducting \RbFeSeSe~could be pinned down to $T_{met}=90$~K and the observation of a gap-like feature at $T_{gap}=61$~K above the superconducting temperature $T_{c}=32$~K indicates the importance of orbital dependent correlations for understanding the involved superconducting pairing mechanism.\cite{Guo10,Tsurkan11,Wang14} Anomalies at these temperatures have been confirmed by pump-probe spectroscopy and Hall measurements.\cite{Li14,Ding14}

Two possible paths to change the correlation strength of the $d_{xy}$ quasiparticles are the application of hydrostatic pressure\cite{Gao2014} and the use of chemical pressure.
The isoelectronic substitution of selenium with sulfur leads to the reduction of the superconducting transition temperature and the upper critical fields in K$_{1-x}$Fe$_{2-y}$Se$_{2-z}$S$_z$.\cite{Lei11,Fabris14}
With respect to the observed hierarchy of temperatures $T_{met}=90$~K, $T_{gap}=61$~K and $T_{c}=32$~K, the evolution of these temperature scales with varying sulfur content will provide important information on how orbital-selective correlation effects influence the onset of superconductivity.

In this work, we perform terahertz (THz) time-domain spectroscopy on
single-crystalline Rb$_{0.75}$Fe$_{1.6}$Se$_{2-z}$S$_z$ that are superconducting
for $z=0$, 0.25, 0.5, and 1.0, and non-superconducting but metallic for $z=1.4$.
We observe a clear increase of the orbital-selective metal-insulator transition temperature from $T_{met}=90$~K to $170$~K with increasing sulfur doping,
showing that correlation effects for the $d_{xy}$ quasiparticles can be tuned by chemical pressure.
In the $z\geq0.25$ systems, the preformed pairs above $T_c$ for $z=0$ are suppressed and the opening of an electronic gap coincides with the onset of superconductivity.
A phase diagram with orbital-selective Mott, metallic, superconducting, and pseudogap-like phases is established as a function of sulfur substitution.

\begin{figure*}[t]
\centering
\includegraphics[width=170mm,clip]{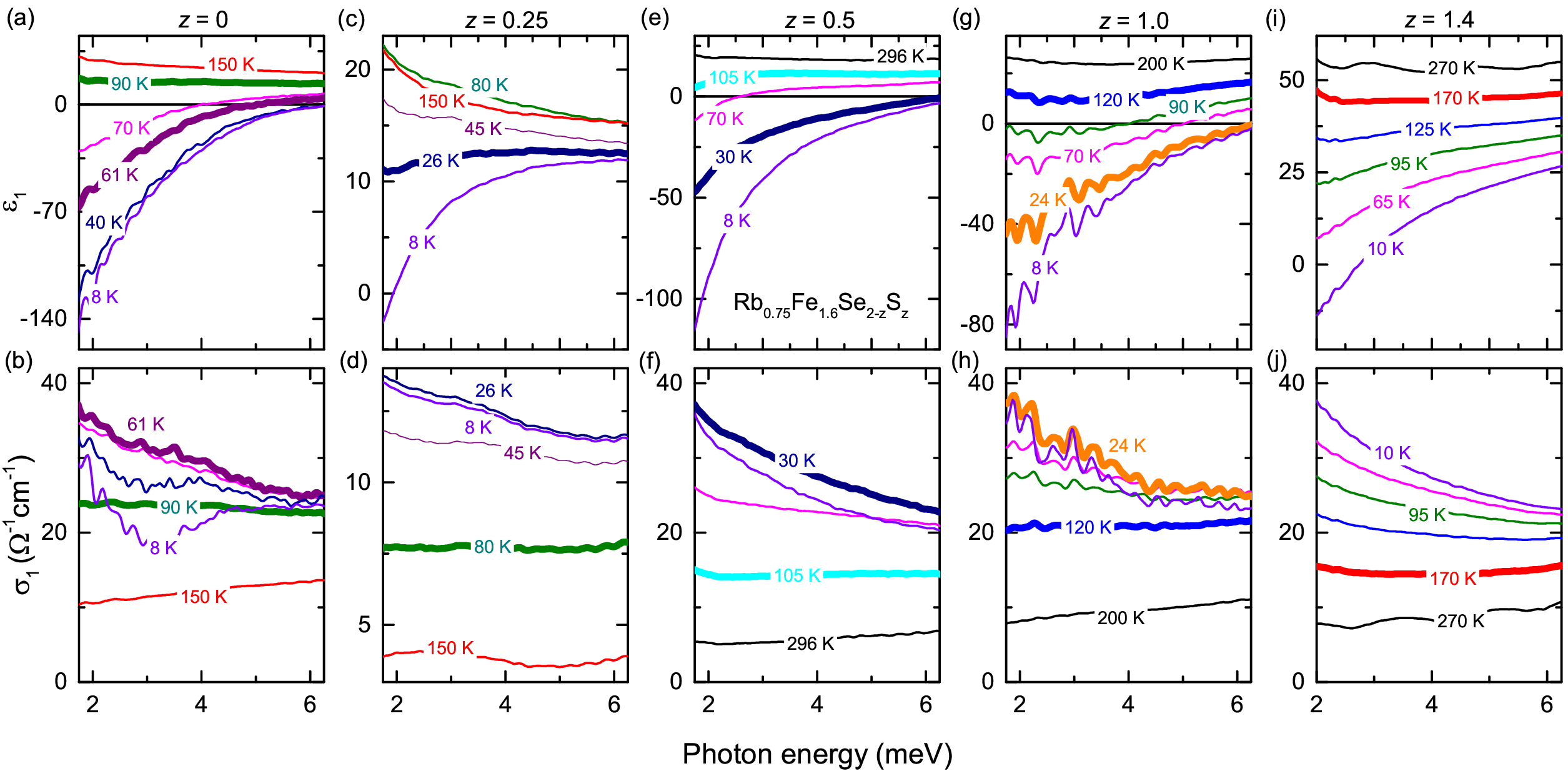}
\vspace{2mm} \caption[]{\label{Fig:epsilon1sigma1} (Color
online)  Dielectric constant $\varepsilon_1$ and optical conductivity $\sigma_1$ of Rb$_{0.75}$Fe$_{1.6}$Se$_{2-z}$S$_z$
for superconductors with
(a)(b) $z=0$,\cite{Wang14}
(c)(d) $z=0.25$,
(e)(f) $z=0.5$, and
(g)(h) $z=1.0$, and metal with
(i)(j) $z=1.4$ as a function of photon energy at various temperatures.
The spectra corresponding to characteristic temperatures $T_{met}$ and $T_{gap}$ are highlighted (see text).
}
\end{figure*}

\section{Experimental details}

Single crystals of the iron chalcogenides were grown
using a Bridgman method following the approach reported in Ref.~\onlinecite{Tsurkan11}.
The phase composition close to Rb$_{0.75}$Fe$_{1.6}$Se$_{2-z}$S$_z$ was determined by wave-length dispersive electron-probe microanalysis.\cite{Tsurkan11}
The superconducting transition temperature $T_c=32$, 24, 28, and 22~K for $z=0$, 0.25, 0.5, and 1.0, respectively,
is determined by measurements of \emph{dc} resistivity and magnetic susceptibility.
The ratios of metallic to semiconducting phases for $z>0$ are comparable with the $z=0$ system. \cite{Ksenofontov11,Ksenofontov,Texier12}
The optical response of the semiconducting phase is nearly independent on temperature or frequency in the THz spectral range.\cite{Charnukha12b}
The details of crystal growth and characterization will be published elsewhere.
The single crystals for optical measurements were prepared with the typical thickness of 40~$\mu$m and cross section of 5~mm$^2$.
Time-domain THz transmission measurements were carried out with
the THz electric field parallel to the crystallographic \emph{ab}-plane using a TPS spectra 3000 spectrometer (TeraView, Ltd.).
A $^4$He-flow magneto-optical cryostat (Oxford Instruments) was used to reach the temperature range from 8 to 300~K.
Transmission and phase shift were obtained from the Fourier transformation of the time-domain signals.
The dielectric constant and optical conductivity were calculated from the transmission and phase shift by modeling the sample as a dielectric slab.\cite{Wang14,Wang15}

\section{Experimental results}

\begin{figure*}[t]
\centering
\includegraphics[width=170mm,clip]{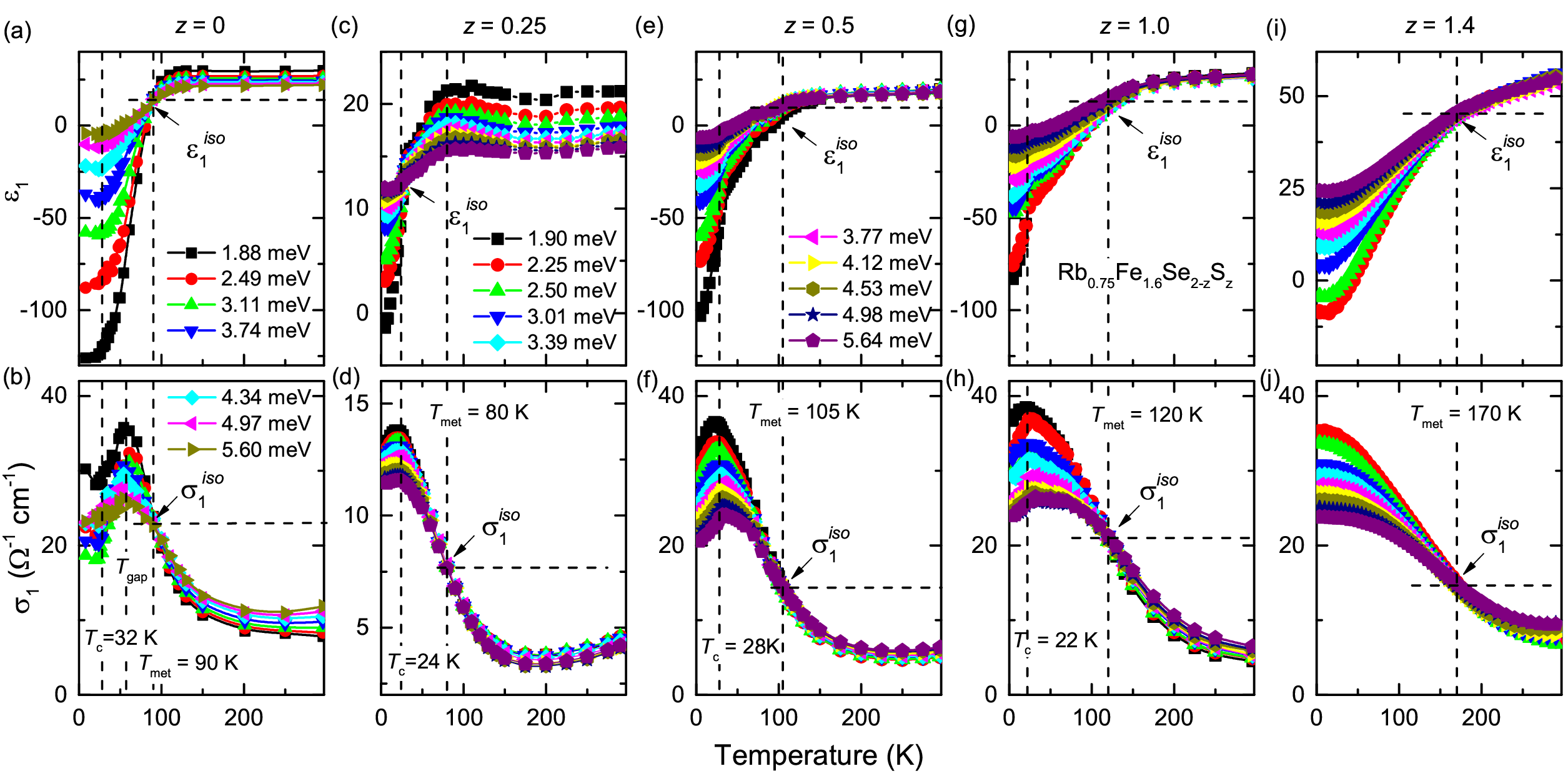}
\vspace{2mm} \caption[]{\label{Fig:SeS_isosbestic} (Color
online) Dielectric constant $\varepsilon_1$ and optical conductivity $\sigma_1$ of Rb$_{0.75}$Fe$_{1.6}$Se$_{2-z}$S$_z$
for superconductors with
(a)(b) $z=0$,\cite{Wang14}
(c)(d) $z=0.25$,
(e)(f) $z=0.5$, and
(g)(h) $z=1.0$, and metal with
(i)(j) $z=1.4$
as a function of temperature for various photon energies.
The orbital-selective metal-insulator transition temperature $T_{met}$ and the superconducting transition temperature $T_c$ are indicated by dashed lines.
$\varepsilon_1^{iso}$ and $\sigma_1^{iso}$ denote the isosbestic points where the dielectric-constant and optical-conductivity curves of different frequencies intersect.}
\end{figure*}

The dielectric constant $\varepsilon_1$
and optical conductivity $\sigma_1$ of Rb$_{0.75}$Fe$_{1.6}$Se$_{2-z}$S$_z$ are shown in Fig.~\ref{Fig:epsilon1sigma1}
for the superconductors with $z=0$, $0.25$, $0.5$, and $1.0$ and the metal with $z=1.4$ as a function of phonon energy at various temperatures.
In the sample with $z=0$ the dominant semiconducting behavior at room temperature changes with decreasing temperature to a metallic response below $T_{met}=90$~K.\cite{Wang14}
$T_{met}$ is defined as the temperature at which an isosbestic point is
emergent in the temperature dependence of the optical conductivity $\sigma_1$ [see Fig.~\ref{Fig:SeS_isosbestic}(b)].
Crossing the isosbestic point from above, $\sigma_1$ increases strongly in the whole spectral range.
Below $T_{met}$, the optical conductivity $\sigma_1$ exhibits Drude-like increase towards lower frequencies [Fig.~\ref{Fig:epsilon1sigma1}(b)].
The $\sigma_1$ spectra of the samples with different sulfur substitutions
follow the same scheme and the values of $T_{met}=80$, 105, 120, and 170~K can be determined for $z=0.25$, 0.5, 1.0, and 1.4, respectively [see Fig.~\ref{Fig:epsilon1sigma1}(d)(f)(h)(j) and Fig.~\ref{Fig:SeS_isosbestic}(d)(f)(h)(j)].

Below the respective $T_{met}$, the dielectric constant $\varepsilon_1$ of the superconductors with $z=0$, 0.5, and 1.0 becomes negative [Fig.~\ref{Fig:epsilon1sigma1}(a)(e)(g)].
For the $z=0.25$ compound, the dielectric constant remains positive in the whole frequency range expect for the lowest temperature and the optical conductivity is relatively low [Fig.~\ref{Fig:epsilon1sigma1}(c)(d)],
although the dc resistivity of the sample behaves similar as for the other doping levels.
In the metallic sample with $z=1.4$, the dielectric constant $\varepsilon_1$ is larger than in the superconducting samples and $\varepsilon_1$ becomes negative only below 10~K [Fig.~\ref{Fig:epsilon1sigma1}(i)],
while the optical conductivity $\sigma_1$ [Fig.~\ref{Fig:epsilon1sigma1}(j)] reaches similar values as in the systems with $z=0$, 0.5, and 1.0.

Figure~\ref{Fig:SeS_isosbestic} shows the dielectric constant $\varepsilon_1$
and optical conductivity $\sigma_1$ of Rb$_{0.75}$Fe$_{1.6}$Se$_{2-z}$S$_z$
as a function of temperature for various photon energies.
For $z=0$, $0.5$, $1.0$, and $1.4$, the dielectric constant is positive and does not show strong temperature or frequency dependence at high temperatures [Fig.~\ref{Fig:SeS_isosbestic}(a)(e)(g)(i)].
On approaching $T_{met}$
from above, the dielectric constant decreases, develops a significant frequency dependence, and becomes negative for $z=0$, $0.5$, and $1.0$
as expected for a coherent metallic response.
A clear kink-like anomaly indicates the onset of superconductivity below $T_c$
which becomes stronger at lower photon energies.
These temperature-dependent features are also reflected by the optical conductivity [Fig.~\ref{Fig:SeS_isosbestic}(b)(d)(f)(h)(j)]:
A gradual and almost frequency-independent increase from room temperature down to the respective $T_{met}$,
where a sharp isobestic point is clearly visible.
The isosbestic point is followed by a frequency-dependent increase down to the gap-formation temperature $T_{gap}$ of 61, 24, 28, and 22~K for the superconducting samples $z=0$, 0.25, 0.5, and 1.0, respectively, and to the lowest temperature for the metallic one ($z=1.4$).
Below $T_{gap}$ the appearance of the preformed gap ($z=0$) or superconducting gap ($z=0.25$, 0.5, and 1.0)
leads to a pronounced maximum followed by a decrease of the optical conductivity to the lowest temperature.

In the case of 1/8 sulfur doping ($z=0.25$), the isosbestic point of the dielectric constant is obtained at the superconducting transition
temperature $T_c$ [Fig.~\ref{Fig:SeS_isosbestic}(c)], while at $T_{met}$, where the optical-conductivity curves intersect [Fig.~\ref{Fig:SeS_isosbestic}(d)], the dielectric constant exhibits a maximum.
Therefore, we consider both temperatures as related to the orbital-selective Mott scenario.

\begin{figure*}[t]
\centering
\includegraphics[width=170mm,clip]{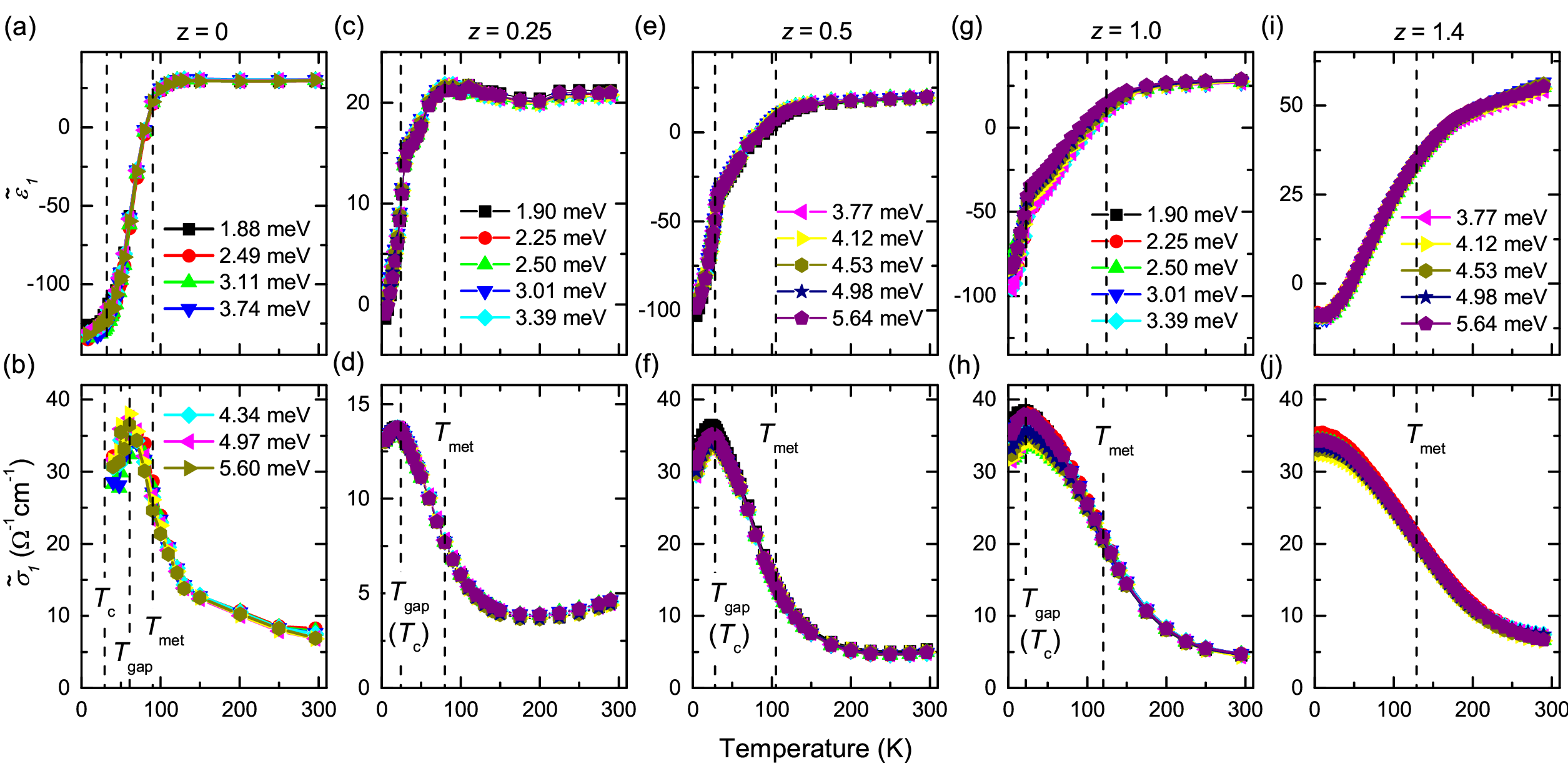}
\vspace{2mm} \caption[]{\label{Fig:SeS_UniversalCurves} (Color
online) Scaled dielectric constant $\widetilde{\varepsilon}_1$ and optical conductivity $\widetilde{\sigma}_1$ of Rb$_{0.75}$Fe$_{2-y}$Se$_{2-z}$S$_z$
for superconductors with
(a)(b) $z=0$,\cite{Wang14}
(c)(d) $z=0.25$,
(e)(f) $z=0.5$, and
(g)(h) $z=1.0$, and metal with
(i)(j) $z=1.4$. Characteristic temperatures are indicated by the vertical lines (see text).}
\end{figure*}

\section{Discussion}

Emergence of an isosbestic point is observed in various correlated systems.\cite{Vollhardt97,Greger13}
In the vicinity of the isosbestic points, one can parameterize the frequency dependence and extract the characteristic temperature-dependent features that reflect the electronic correlations in the systems.\cite{Greger13}
At the isosbestic points, the dielectric constant $\varepsilon_1$ and optical conductivity $\sigma_1$ are frequency independent and the frequency dependence of $\varepsilon_1$ and $\sigma_1$
can be approximated in the vicinity of the isosbestic points by
\begin{align}\label{Eq:Isos_1}
\varepsilon_1(T,1/\omega)&=\varepsilon_1(T,1/\omega_0)+(1/\omega-1/\omega_0)E_1(T),\\
\label{Eq:Isos_2} \sigma_1(T,\omega)&=\sigma_1(T,\omega_0)+(\omega-\omega_0)S_1(T),
\end{align}
where the parameters
\begin{align}\label{Eq:Isos_fittingParam}
E_1(T)&=\frac{\partial \varepsilon_1(T,1/\omega)}{\partial (1/\omega)}\bigg |_{1/\omega=1/\omega_0},\\
S_1(T)&=\frac{\partial \sigma_1(T,\omega)}{\partial \omega}\bigg |_{\omega=\omega_0},
\end{align}
can be obtained by fitting Eqs.~(\ref{Eq:Isos_1}) and (\ref{Eq:Isos_2}) to the experimental results.\cite{Vollhardt97,Greger13,Wang14}
Using the resulting scaling relations
\begin{align}\label{Eq:Isos_mastercurve1}
\tilde{\varepsilon}_1(T,1/\omega_i)&=\varepsilon_1(T,1/\omega_i)-(1/\omega_i-1/\omega_0)E_1(T),\\
\label{Eq:Isos_mastercurve2} \tilde{\sigma}_1(T,\omega_i)&=\sigma_1(T,\omega_i)-(\omega_i-\omega_0)S_1(T).
\end{align}
for the dielectric constant $\varepsilon_1(T,1/\omega)$ and optical conductivity $\sigma_1(T,\omega)$,  the experimental curves $\tilde{\varepsilon}_1(T,1/\omega_i)$ and $\tilde{\sigma}_1(T,\omega_i)$ for different frequencies ($\omega_i$) are expected to collapse on a single master curve for the dielectric constant and optical conductivity, respectively.\cite{Greger13}

According to the Eqs.~(\ref{Eq:Isos_1}) and (\ref{Eq:Isos_2}), the fittings are performed for the dielectric constant and the optical conductivity,
respectively, for the iron-chalcogenide superconductors Rb$_{0.75}$Fe$_{1.6}$Se$_{2-z}$S$_z$ with $z=0$, 0.25, 0.5, and 1.0, and the metal with $z=1.4$.
The obtained results according to Eqs.~(\ref{Eq:Isos_mastercurve1}) and (\ref{Eq:Isos_mastercurve2}) are shown in Fig.~\ref{Fig:SeS_UniversalCurves}.
For each doping level, the curves of different frequencies collapse onto that with the lowest frequency.
The obtained master curves confirm the validity of the parameterizations and the characteristic temperatures determined by the isosbestic points as highlighted in Fig.~\ref{Fig:epsilon1sigma1} and Fig.~\ref{Fig:SeS_isosbestic}.

In the $z=0$ system, the emergence of a metallic optical response below $T_{met}=90$~K
was attributed to an orbital-selective Mott transition,\cite{Wang14}
where the $d_{xy}$ band contributes to the metallic properties only below $T_{met}$,
while the $d_{xz}/d_{yz}$ bands retain their metallic features both below and above $T_{met}$.\cite{Yi13,Yu13}
Incoherent hopping process via the $d_{xz}/d_{yz}$ orbitals
at the interface of the superconducting and the antiferromagnetic phases
can lead to large scattering rates in the optical response of the corresponding quasiparticles \cite{Huang13,Wang14}.
In contrast, the $d_{xy}$ channel remains almost unaffected by the proximity effect and
reveals its metallic optical response at low frequencies via its larger mass normalization.\cite{Yi13,Wang14}

This scenario also can be applied to the metal-to-insulator transition observed for the sulfur-doped systems, given their similarities with the $z=0$ compound in band structure,\cite{Feng15,Yi15b} electronic valence state, and optical properties.
However, the increase of the metal-to-insulator transition temperature for $z>0.25$ with increasing sulfur doping indicates
that the mass renormalization for the $d_{xy}$ band is lowered, and thus the reduction of the electron correlations.
Following the isoelectronic scenario, our results can be compared to a recent theoretical study,
where it was shown that $T_{met}$ increases with decreasing intra-orbital Coulomb repulsion.\cite{Yi15}

The orbital-selective metal-insulator transition temperature $T_{met}$ determined by the THz spectroscopy is clearly distinct from the temperature where a broad maximum is observed in the temperature-dependent \emph{dc} resistivity. \cite{Lei11}
The latter can be described as a superposition of a metallic and a semiconducting contribution according to the phase-separated nature of the samples, and thus strongly depends on volume fraction and geometry of the metallic phase. \cite{Shoemaker12}
With increasing sulfur content, $T_{met}$ increases monotonically,
while a non-monotonic change of the temperature of resistivity maximum was observed in K$_{x}$Fe$_{2-y}$Se$_{2-z}$S$_z$.\cite{Lei11}
Since the $d_{xz}$/$d_{yz}$ quasiparticles retain their metallic contribution to the \emph{dc} conductivity,\cite{Wang14} $T_{met}$ is located within the metallic regime and no anomaly is observed in the resistivity at $T_{met}$.

The formation of a superconducting gap is usually reflected by the suppression of optical conductivity,
since the spectral weight at finite frequency is transferred to the superconducting condensate at zero frequency.
The frequency, at which the minimum of optical conductivity occurs, usually provides an estimate for the superconducting gap $2\Delta$.\cite{Mattis58,Dressel02,Homes12R,Yuan12,Bernhard12}
In the $z=0$ system, the suppression of optical conductivity occurs at $T_{gap}=61$~K much higher than $T_c=32$~K [Fig.~\ref{Fig:epsilon1sigma1}(b)].
A smaller gap of $2\Delta=3.2$~meV at 8~K can be revealed from the observed suppression of optical conductivity,\cite{Wang14}
in addition to a larger superconducting gap at $2\Delta\sim16$\,--\,20~meV that was resolved by ARPES.\cite{Xu12,Borisenko12}
The preformed gap associated with the quasiparticles in the $d_{xy}$ band
is not observed in the sulfur-doped superconductors.
In the $z\geq0.25$ samples,
the suppression of optical conductivity occurs only below $T_c$.
While the difference in optical conductivity below and above $T_c$ is very small for $z=0.25$ but with a broad minimum at $2\Delta=2.4$~meV [Fig.~\ref{Fig:epsilon1sigma1}(d)],
it is almost frequency independent for $z=0.5$ and $z=1.0$ in the investigated spectral range [Fig.~\ref{Fig:epsilon1sigma1}(f)(h)].
Hence in the $z=0.5$ and $1.0$ samples the smaller gap is already suppressed.
The constant suppression of the optical conductivity below $T_c$ is a result of the opening of a superconducting gap
whose energy is out of the investigated frequency range.

\begin{figure}[t]
\centering
\includegraphics[width=80mm,clip]{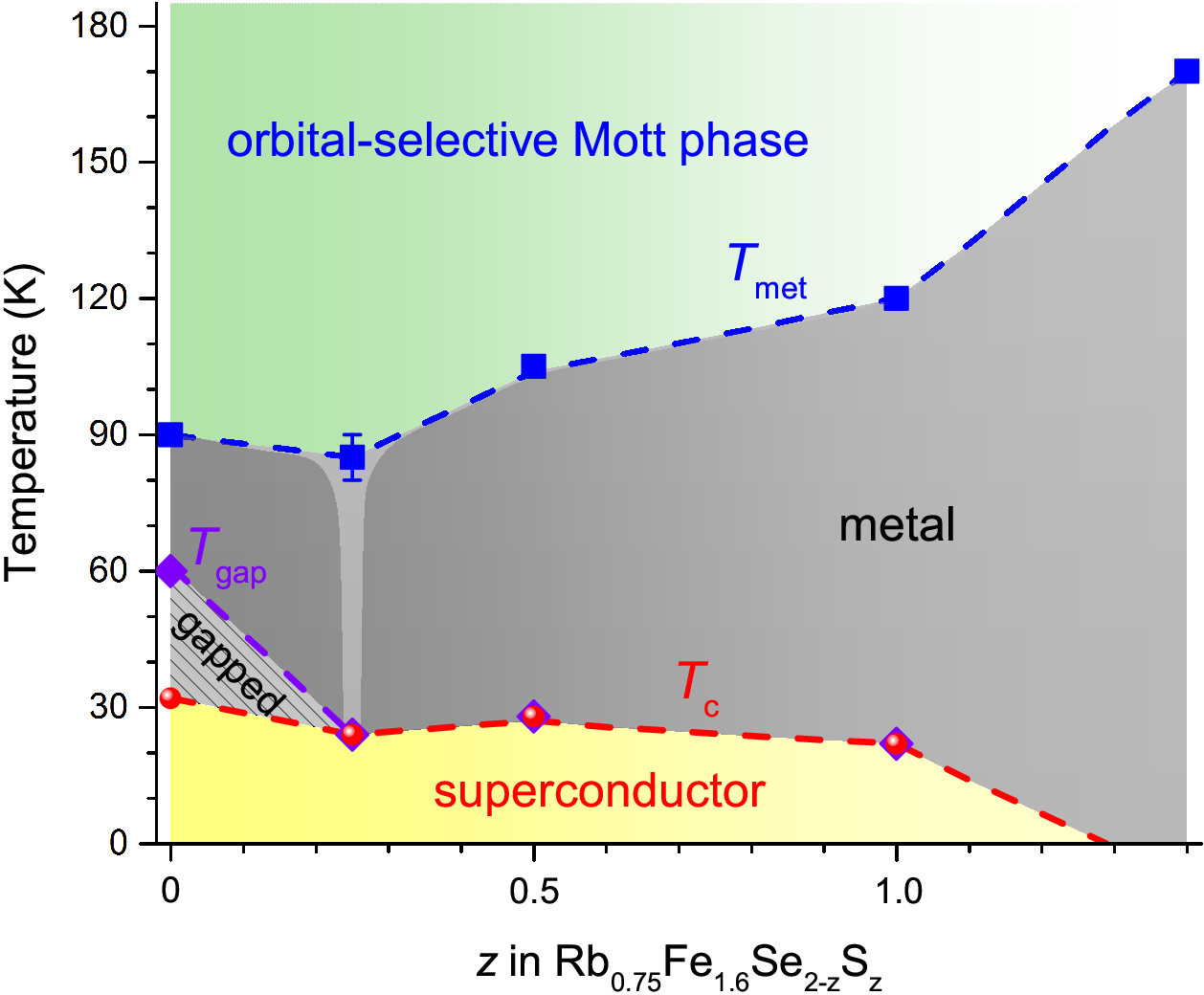}
\vspace{2mm} \caption[]{\label{Fig:SeS_PhaseDiagramm} (Color
online) Phase diagram with orbital-selective Mott, metallic, preformed-gap, and superconducting phase for Rb$_{0.75}$Fe$_{1.6}$Se$_{2-z}$S$_z$.
$T_{gap}$ and $T_{met}$ are obtained from the terahertz spectroscopy.
$T_c$ is determined by \emph{dc} resistivity and magnetic susceptibility measurements.
}
\end{figure}

The observation of the metal-to-insulator, gap formation, and superconducting transition temperatures
as a function of sulfur content
enables us to establish a phase diagram with orbital-selective Mott, metallic, preformed-gap, and superconducting phases, as displayed in Fig.~\ref{Fig:SeS_PhaseDiagramm}.
For the $z=0.25$ system with a local minimum of $T_c$, the separation of the isosbestic points in the optical conductivity and dielectric constant [Fig.~\ref{Fig:SeS_isosbestic}(c)(d)] is also indicated in the phase diagram.
These observations may be
interpreted in terms of stabilization or enhancement of electron correlations in the $d_{xy}$ orbital channel.
The origin remains unclear but additional ordering effects
associated with the 1/8 sulfur doping level ($z=0.25$) may be possible ingredients for such behaviors.\cite{Geck04}
At higher doping levels, the monotonic decrease of electron correlations and superconducting transition temperature is restored.

In summary, using terahertz spectroscopy we have investigated the orbital-selective metal-insulator transition associated with quasiparticles with $d_{xy}$ orbital character in the iron-chalcogenide superconductors Rb$_{0.75}$Fe$_{1.6}$Se$_{2-z}$S$_z$ with $z=0.25$, 0.5, and 1.0, and the metal with $z=1.4$. In comparison to $T_{met}=90$~K in the undoped system, the orbital-selective metal-insulator transition temperature
is strongly increased with increasing sulfur substitution up to 170~K for $z=1.4$,
while the superconducting transition is reduced and finally suppressed.
This observation is a clear indication that the electron correlations in the $d_{xy}$ orbital channel are reduced by the isoelectronic substitution.
Varying the correlation strength of the $d_{xy}$ orbital channel in Rb$_{0.75}$Fe$_{1.6}$Se$_{2-z}$S$_z$ can be regarded as an efficient way to tune the pairing in the $d_{xy}$ channel.
We believe that the suppression of the pseudo-gap like feature is related to reduced spin fluctuations of the $d_{xy}$ quasiparticles.

\begin{acknowledgments}
We acknowledge partial support by the Deutsche Forschungsgemeinschaft via the Transregional Research Collaboration TRR 80: From Electronic Correlations to Functionality (Augsburg - Munich - Stuttgart), within the SPP~1458, and by the Project DE 1762/2-1.
Z.W. acknowledges support by the Chinesisch-Deutsches Zentrum f\"{u}r Wissenschaftsf\"{o}rderung and hospitality of the International Center for Quantum Materials at Peking University.
\end{acknowledgments}


\end{document}